\pdfoutput=1
\RequirePackage{ifpdf}
\ifpdf 
\documentclass[pdftex]{sigma}
\else
\documentclass{sigma}
\fi

\numberwithin{equation}{section}

\newtheorem*{Theorem*}{Theorem}

 { \theoremstyle{definition}

 }

\begin{document}

\allowdisplaybreaks

\newcommand{\arXivNumber}{2307.15222}

\renewcommand{\PaperNumber}{099}

\FirstPageHeading

\ShortArticleName{Newton's Off-Center Circular Orbits and the Magnetic Monopole}

\ArticleName{Newton's Off-Center Circular Orbits\\ and the Magnetic Monopole}

\Author{Dipesh BHANDARI~$^{\rm a}$ and Michael CRESCIMANNO~$^{\rm b}$}

\AuthorNameForHeading{D.~Bhandari and M.~Crescimanno}

\Address{$^{\rm a)}$~Department of Physics and Astronomy, Texas A\&M University-Commerce, TX, 75429, USA}
\EmailD{\href{mailto:dbhandari@leomail.tamuc.edu}{dbhandari@leomail.tamuc.edu}}

\Address{$^{\rm b)}$~Department of Physics and Astronomy, Youngstown State University,\\
\hphantom{$^{\rm b)}$}~Youngstown, OH, 44555, USA}
\EmailD{\href{mailto:dcphtn@gmail.com}{dcphtn@gmail.com}}

\ArticleDates{Received July 31, 2023; in final form December 08, 2023; Published online December 17, 2023}

\Abstract{Introducing a radially dependent magnetic field into Newton's off-center circular orbits potential so as to preserve the $E=0$ dynamical symmetry leads to a unique choice of field that can be identified as the inclusion of a magnetic monopole in the inverse stereographically projected problem. One finds also a phenomenological correspondence with that of the linearly damped Kepler model. The presence of the monopole field deforms the symmetry algebra by a central extension, and the quantum mechanical version of this algebra reveals a number of zero modes equal to that counted using the index theorem of elliptic operators.}


\Classification{37J37; 19K56}

\section{Motivation}
In Book~1, Proposition~7, Problem~2 of his 1687 Philosophiae Naturalis Principia Mathematica, Isaac Newton finds the unique smooth, radially symmetric potential whose force results in off-center circular orbits (non-relativistic)~\cite{newton}. There are several remarkable features of these circular orbits in this potential. Their orbital axis (length and angle) implies, just as in the Kepler case, the existence of two constants of motion in addition to the angular momentum, as was detailed in~\cite{olshanii23}. The Poisson algebra of these constants of motion closes on the $E=0$ strata into an $\mathfrak{so}(3)$ algebra, as in that of the Runge--Lenz symmetry in the Kepler problem.

Here we explore a deeper algebraic and geometric connection between the Kepler case and the~$E=0$ strata of Newton's off-center circular orbit problem by first, as motivation, recalling that Kepler orbits subject to frictional force proportional to the velocity evolve in $E$ while preserving their orbital eccentricity and orientation \cite{hamilton08}, a consequence of the `rigidity' of the $\mathfrak{so}(3)$ Runge--Lenz symmetry (note this is also the case for the $\mathfrak{so}(3)$ symmetry in the isotropic 2-d harmonic oscillator with frictional forces linear in the velocity, a system also directly relatable to the Kepler problem~\cite{faure}). In seeking to test an analogue of this `rigidity' in the off-center circular orbit problem of Proposition 7, Problem 2 of the Principia, a frictional force proportional to the velocity will not be appropriate since it will evolve into $E \ne 0$ orbits for which there is no $\mathfrak{so}(3)$-like symmetry. Giving the particles charge and subjecting them to a magnetic forces (also linear in the velocity) on the other hand, will in general also modify the spectrum, but below we find a unique choice of magnetic field that again results in an $\mathfrak{so}(3)$-like symmetry on $E=0$ orbits.\looseness=-1

In the Kepler problem, adding a magnetic field generally breaks the $\mathfrak{so}(3)$ algebra, as all orbits except concentric circular orbits now precess and the circular ones have $E < 0$. Adding a magnetic field to the off-center circular orbit Newton problem has a markedly different effect. Below we show that there is a unique magnetic field with that potential for which the Poisson algebra of first integrals of motion for $E=0$ satisfy an $\mathfrak{so}(3)$-like algebra. That algebra turns out to be a central extension of that found in~\cite{olshanii23}, again leading to off-center circular orbits.

The stereographic projection of the original Newton problem onto a sphere of radius ${\cal R}$ maps the $E=0$ off-center orbits into geodesics on that sphere \cite{olshanii23}. In the magnetized problem as we frame it below, the projection of the $E=0$ orbits map onto the intersection of that sphere with a plane not going through the origin. These orbits are those of a charge confined to the sphere in which a magnetic monopole occupies its center. The charge of the monopole is exactly twice the central extension of the Poisson algebra.

\section{Algebra and dynamics}
We consider all orbits as solutions of the equations of motion on the two-dimensional plane and all vectors (denoted in bold) as two-dimensional unless explicitly indicated.

To fix notation, let the Hamiltonian of a non-relativistic particle of mass $m=1$ in two-dimensions with momentum $\mathbf{p}$ be written as
\begin{equation}
H = \frac{\mathbf{p}^2}{2}+V(r).
\label{eq1}
\end{equation}
Throughout $r$, $\phi$ will refer to the usual polar co-ordinates on the plane, and $x=r\cos\phi$, ${y=r\sin\phi}$ the cartesian co-ordinates.
Following Newton \cite{newton}, in \cite{olshanii23} for $B=0$ (no magnetic field), for the unique choice of \smash{$V(r) = -\frac{\alpha}{(r^2+{\cal R}^2)^2}$} with a constant ${\cal R}$,
one has in addition~to \smash{${\tilde L}_z = xp_y-yp_x$}, the two-dimensional vector \smash{$\mathbf{I} = {\tilde L}_z \mathbf{r} + (\mathbf{r}\cdot \mathbf{p})\mathbf{e}_z\times \mathbf{r} + {\cal R}^2 \mathbf{e}_z\times \mathbf{p}$} generating an enveloping algebra that commutes with $H$ on the $H=0$ strata, where $\mathbf{e}_z = \mathbf{e}_x\times \mathbf{e}_y$. To recapitulate, as shown in \cite{olshanii23} for $B=0$, we have
\begin{equation}
\{\mathbf{I},H\} = -4H \mathbf{e}_z\times \mathbf{r}, \qquad \big\{{\tilde L}_z,H\big\} =0,\qquad \big\{\mathbf{I}, {\tilde L}_z\big\}=\mathbf{e}_z\times \mathbf{I},\qquad \{I_x, I_y\} = 4{\cal R}^2 {\tilde L}_z.
\label{B0}
\end{equation}
This observation greatly simplifies the determination of the orbits at $H=0$. For $H=0$ the $I$ are constants of motion and dividing them by ${\tilde L}_z = r^2 \frac{{\rm d}\phi}{{\rm d}t}$ gives two first order, linear equations in $r(\phi)$ and $\frac{{\rm d} r}{{\rm d}\phi}$. Eliminating the $\frac{{\rm d} r}{{\rm d}\phi}$ gives an $r(\phi)$, an off-center circle whose radius $R$ squared is half the quadrature sum of \smash{$I_x/{\tilde L}_z$} and \smash{$I_y/{\tilde L}_z$} and whose angular orientation is given in terms of $\arctan(I_y/I_x)$ \big(the distance between the orbital center and the origin being $\sqrt{R^2-{\cal R}^2}\big)$.

For $B\ne 0$, we use minimal substitution, $\mathbf{P} = \mathbf{p}+\mathbf{A}$, where $\mathbf{p}$ is the conjugate momentum vector and $\mathbf{A}$ is the two-component vector potential, and write the Hamiltonian as
\begin{equation}
H = \frac{\mathbf{P}^2}{2}+V(r)
\label{eqB}
\end{equation}
with the foregoing $V$. We define $B$, the magnetic field, here a pseudoscalar density on the two-plane, through,
\begin{equation}
\{P_x, P_y\} = B.
\label{eq2}
\end{equation}
We limit our analysis to the case where $B$ depends only on $r$, and define $G = 2\int B(r) r {\rm d}r$.
The magnetic generalization of ${\tilde L}_z$ we denote $L_z = xP_y-yP_x + G/2$, which although it clearly does not commute with the Hamiltonian in \eqref{eq1} it does with the one in \eqref{eqB}, that is, $\{L_z, H\}=0$. The convenient magnetic generalization of the vector $\mathbf{I}$ we call $\mathbf{J}$ and define
\begin{equation}
\mathbf{J} = \left(L_z+\frac{G}{2}\right) \mathbf{r} + (\mathbf{r}\cdot \mathbf{P})\mathbf{e}_z\times \mathbf{r} + {\cal R}^2 \mathbf{e}_z\times \mathbf{P}.
\label{J}
\end{equation}
Requiring that $\{\mathbf{J}, H\}=-4H \mathbf{e}_z\times \mathbf{r}$ forces the magnetic field to have the unique solution $B(r) = -Q/\big(r^2+{\cal R}^2\big)^2$ for some constant $Q$. Another brief calculation indicates
\begin{gather}
\{\mathbf{J}, H\}=-4H \mathbf{e}_z\times \mathbf{r}, \qquad \{L_z,H\}=0,\qquad \{\mathbf{J},L_z\} = \mathbf{e}_z\times \mathbf{J},\nonumber\\
\{J_x, J_y\} = 4{\cal R}^2 L_z - Q,
\label{Bnot0}
\end{gather}
which is a central extension (by $Q$) of the $B=0$ case in \eqref{B0}.
This algebra again indicates that the $L_z$, $\mathbf{J}$ are constants of the motion for $H= E = 0$ orbits. As in the $B=0$ case, we use these constants to find the shape of the $E=0$ orbits. Resolving the expressions $\mathbf{J}\cdot \mathbf{r}/L_z$ and $\mathbf{J}\times \mathbf{r}/L_z$ we find the periodic $H=0$ orbits satisfy
\begin{equation}
\frac{\mathbf{J}\cdot \mathbf{r}}{L_z} = r^2 - {\cal R}^2 + \frac{Q}{2L_z},\qquad  \frac{\mathbf{J}\times \mathbf{r}}{L_z} = -\big(r^2 + {\cal R}^2\big)\frac{(\mathbf{r} \cdot \mathbf{P})}{L_z}.
\label{orbitConstraints}
\end{equation}
Note that the second equation does not involve $Q$. The first equation above is basically the integral of the second one, in which term proportional to $Q$ emerges as a constant of integration. To see that in more detail, without loss of generality, take $\mathbf{J} = (J,0)$ pointing in the ${\mathbf{\hat x}}$ direction only and use the Hamilton equation, $\mathbf{P} = \frac {{\rm d}\mathbf{r}}{{\rm d}t}$. Writing $L_z = r^2 \frac {{\rm d}\phi}{{\rm d}t}$ on the right-hand side while keeping $L_z$ as a constant on the left-right side; using finally \smash{$\mathbf{r} \cdot
\frac {{\rm d}\mathbf{r}}{{\rm d}t} = \frac {{\rm d}r^2}{2{\rm d}t}$} one can then integrate the second equation to get the first. Note that for this choice of $\mathbf{J}$ the first equation is just the equation for a circle centered about $\big(\frac{J}{2L_z}, 0\big)$ of radius squared equal to ${\cal R}^2 +\frac{J^2}{4L_z^2} - \frac{Q}{2L_z}$, requiring the latter to be non-negative.

Varying the choices of $J$ and $Q$ it is clear that for a fixed ${\cal R}$ one sign $(-)$ of $\frac{Q}{2L_z}$ corresponds to circles whose chord through the origin extend beyond $\pm {\cal R}$, while for the other choice of sign~$(+)$ is within $\pm {\cal R}$.

Finally, it is useful to compute the quadratic Casimir element of the Poisson algebra~\eqref{Bnot0},
\begin{equation}
C_2 = \frac{\mathbf{J}\cdot \mathbf{J}}{4{\cal R}^2} + \biggl(L_z-\frac{Q}{4{\cal R}^2}\biggr)^2 = \frac{\big(r^2+{\cal R}^2\big)^2 H}{2{\cal R}^2} +\frac{\alpha}{2{\cal R}^2}+ \frac{Q^2}{16{\cal R}^4}.
\label{Casimir}
\end{equation}
For $E=0$, the $C_2$ must be a positive constant; this limits the magnitude of $|\mathbf{J}|$ and~$L_z$ for each~$Q$. Note also that combining $C_2$ for $E=0$ orbits with the \eqref{orbitConstraints}, one arrives at a simple formula for the radius $R$ (in the plane) of the orbit, $R = \sqrt{\frac{\alpha}{2}} \frac{1}{L_z}$, true for all $Q$. The value of $|J|$ is directly associated with the distance, ${l}$ between the center of the orbit an the origin, \smash{${l}^2 = R^2-{\cal R}^2+\frac{Q}{2L_z}$}. For each fixed $Q$, the convexity of the Casimir element limits the magnitude of the $L_z$, and so this indicates that there is a minimum possible $R$ for that $Q$. It is only in the limit $|Q| \rightarrow \infty$ that the $L_z$ can grow without bound and the orbit can be of zero radius.

Historically, in dealing with a central potential combined with a magnetic monopole, some authors have noted that including an additional potential term proportional to the square of the magnetic charge and varying as $1/r^2$ can restore the dynamical symmetry \cite{cisneros} to what it was before the addition of the magnetic monopole. That is, the gauge covariant integrals of motion now commute with the new Hamiltonian and satisfy the same algebra as it was at $Q=0$. At an algebraic level, the analogous term to add to $H$ of \eqref{eqB} would be $-Q^2/\big[8{\cal R}^2\big(r^2+{\cal R}^2\big)^2\big]$ which preserves the algebra and leads to the same Casimir element since it is just a change in $\alpha$.

A simpler way to say this physically is that adding in a magnetic monopole field in the cases studied in \cite{cisneros} causes the orbits to not close, and the additional potential term $\sim Q^2/r^2$ added to the $H$ makes the orbits again close in those cases. The case of Newton's off-center circular orbit potential is quite different; here the inclusion of a magnetic monopole field preserves orbital closure and the addition of the term analogous to ``$Q^2/r^2$" simply keeps the problem in the original family of Hamiltonians.

For completeness, we have also computed the orbital period for this potential. For $E \ne 0$, the period integral is elliptic functions, but as expected, one of the branchcuts degenerates in the~$E=0$ limit where the period integral reduces to a trigonometric function. One finds for each $E=0$ orbits the period \[T= \frac{\pi\alpha}{|L_z^3|} + \frac{\pi Q}{2L_z^2}.\] For a fixed $Q$ on the physical branch \big(on which the most negative $Q/(2L_z)$ can be is $-R^2+{\cal R}^2$\big) the maximum period occurs for $L_z \rightarrow 0$, $R\rightarrow \infty$. Note finally that the minimum possible period is 0 in the limit of large $|Q/L_z|$ only, as expected from the Larmor limit ($R\rightarrow 0$).

\section{The magnetic monopole via the stereographic projection}
Note that the conformal transformation between the orbital plane and the stereographic representation from \cite{olshanii23} is given in that paper by
\begin{equation}
g_{\mu\nu}^{\rm plane} = \biggl(\frac{r^2 + {\cal R}^2}{2{\cal R}^2}\biggr)^2 g_{\mu\nu}^{S^2}.
\label{conformal}
\end{equation}
Since the field $B$ is defined via the commutators of two momenta, in keeping with \[F_{\mu\nu} = B\epsilon_{\mu\nu} = \{P_\mu, P_\nu\},\] we see that the quantity $B$ transforms as
\[
B^{\rm plane} = \bigl(\frac{2{\cal R}^2}{r^2 + {\cal R}^2}\bigr)^2 B^{S^2},
\]
 and using the form of $B^{\rm plane}$ in~the previous section we learn that the \smash{$B^{S^2} = -\frac{Q}{4{\cal R}^4}$}, a constant, that is, the unique $B$ field that closes the algebra in the plane is a magnetic monopole as viewed in the stereographic sphere, of dimensionless total flux $= 2\pi M = -\frac{\pi Q}{{\cal R}^2}$ (where we define the monopole charge $M$, which is integer in the quantum case with $\hbar=1$).
Note also that as the charge is a two-form integrated over a surface, it is a topological invariant and, so we of course get the very same charge integrating $B^{\rm plane}$ in the area element ${\rm d}x{\rm d}y$ of the plane over the entire plane.

The free motion of charged mass point in three dimensions in the presence of a magnetic monopole is a celebrated classical \cite{golo, shnir} and quantum mechanical problem \cite{silva,grossman, kemp,song} and here, restricting the mass's motion to an $S^2$ with a central monopole breaks the symmetry algebra to just $\mathfrak{so}(3)$. The $\mathfrak{so}(3)$ invariance of that problem indicates that geodesic motion in such a~case would of course have three independent constants of motion. In our problem the relaxed constraint that the algebra only commute with the Hamiltonian on the single strata~$H=0$ allows us to evade the classification of superintegrable systems of this type in the 2-d plane \cite{winternitz3}.

Under stereographic projection circles map to circles, and the three constants of motion can be thought of as two moduli for the center of the circle and one for the radius (on either the plane or the sphere). The radius of the orbit on the sphere can be conveniently written in terms of the half angle $\gamma$ of the cone whose apex is the center of the sphere. For a unit charge moving uniformly in that orbit, by elementary consideration of Newton's equation we find its angular velocity would be $\omega = \frac{|Q|}{4m{\cal{ R}}^4\cos\gamma}$ when $\gamma \ne \pi/2$. Note that the period of the orbit in the plane and the period of a free charge bound to a sphere (see above) are quite different, indicating that the stereographic projection relates a geometric similarity between the two problems but is not a~dynamical equivalence. We do however note that the sign of $Q/L_z$ does indicate whether~$\gamma<\pi/2$ or $\gamma > \pi/2$, as the sign at large $Q/L_z$ can localize orbits outside of $r={\cal R}$ or inside that circle in the plane. Finally, unlike the case with a charge moving on a~plane with constant $B$ field, due to the curvature (sphere)/potential(plane) there is no simple relation between the angular momentum of the orbit and the magnitude of the magnetic flux it encompasses.

\begin{figure}[t]
\centerline{\includegraphics[width=5.5in]{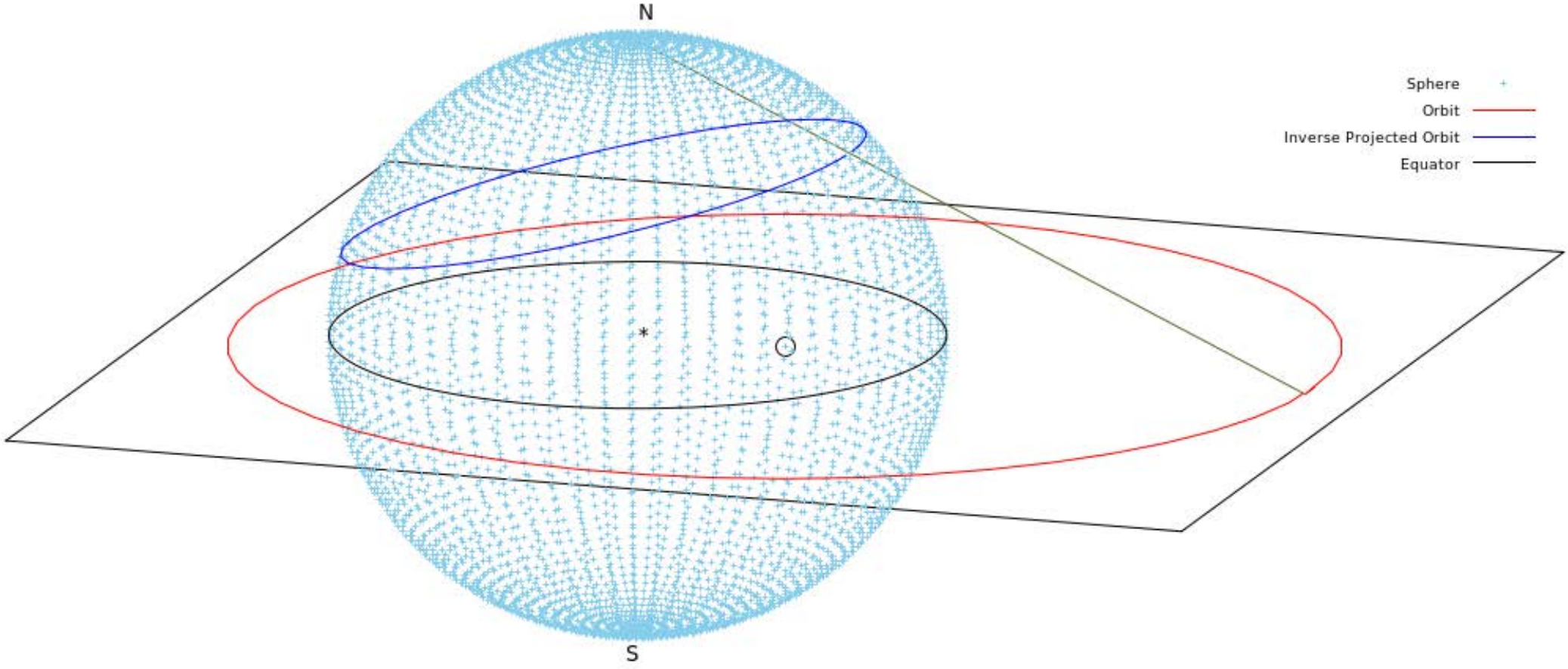}}
\caption{Basic montage of the problem; red orbit and orbital, along with the stereographic sphere (light blue) and the projection (green line) onto the sphere (orbit image is blue). For reference, the equator of the sphere is shown in black with its center at `$\ast$' and the center of the orbit at `$\circ$'. Poles are marked with `N' and `S'.}\label{fig1}
\end{figure}

\section{Discussion}
In a sense, the findings described here are not surprising: the addition of the magnetic field has caused some of the orbits that were precessing to stop precessing (and the ones that were not precessing to start precessing). The spatial form of the field is such that these closed orbits all happen at zero energy. Note that the general solution for the orbital shape at $E\ne 0$ can be given in terms of elliptic functions; they degenerate to trigonometric functions at $E=0$.

From the algebraic point of view, much of the forgoing has also been known since the early study of monopoles by Dirac and Schwinger \cite{dirac}. For example, it is well known that in the presence of a monopole the mechanical angular momentum and geometrical operator that causes rotations about an axis are not the same, but are related by an overall term proportional to $Q{\mathbf{\hat r}}$, as was also used here \cite{grossman,shnir}. Likewise the idea that the magnetic field results in a central extension of the dynamical algebra is strongly reminiscent of the case of a constant field in the plane, the celebrated Landau problem.

It is noteworthy to also compare the linear stability of the orbits. Instead of using the effective potential approach we ask what bound (not closed) orbits nearby are accessible with an arbitrary small change in the energy $E$ and the angular momentum $L_z$. For concreteness, the general bound orbit has an innermost and outermost excursion from the origin; let $a$ represent the outermost distance and $v_a$ the particle's velocity there. We presume that there are other orbits nearby in $a$, $v_a$, so for this family locally we have $E = \frac{1}{2} v_a^2 + V(a)$ and $L_z = av_a$ where $V$ is the potential energy function (mass $m$ = 1). The general excursion $(\delta E,~\delta L_z)$ can be rendered as an excursion in $(\delta a,~\delta v_a)$ as
\begin{equation}
(\delta E,~\delta L_z)^t = {\cal M} (\delta a,~\delta v_a)^t, \qquad
{\cal M} = \biggl[
\begin{matrix}
V' & v_a\cr
v_a & a\cr
\end{matrix}\biggr],
\label{landau}
\end{equation}
where $V' = \frac {{\rm d}V}{{\rm d}r}|_a$, indicating that the obstruction to linear response in orbital parameters $a$, $v_a$ under a general $\delta E$, $\delta L_z$ is $0 = \det ({\cal M}) = \big(aV'-v_a^2\big) = (aV'+2V-2E)$. This relation holds for arbitrary magnetic field. For the \smash{$V(r) = -\frac{\alpha}{(r^2+{\cal R}^2)^2}$} used here, the determinant $\det ({\cal M})$ is zero only at $a={\cal R}$. Since for $E=0$ that is a boundary case (the centered circle: for which $\delta a$ must change sign) this shows that all the off-center circular orbits of the Newton problem for any~$Q$ are linearly stable. Note that for the family of orbits that are circles with centers at the origin (so $E\ne 0$), the same computation gives \smash{$\det ({\cal M}) = \frac{\omega Qa^2}{(a^2+{\cal R}^2)^2}$} where $\omega$ is the orbital frequency, thus in the original Newton problem $(Q=0)$ orbits centered at the origin are linearly unstable. In a direct sense, the existence of the closed Poisson algebra, because of the constants of motion and their convexity due to the Casimir element, forces the orbits (in this case, restricted again to the $H=0$ strata) to be stable, perhaps even beyond the linear perturbative regime.

The foregoing brings us back to our original motivation, to see in this off-center circular potential a generalization of the fact that the Kepler problem/isotropic harmonic oscillator system subject to linear frictional damping preserves two of three orbital constants of motion. Recall that in the Kepler case, linear friction---a non-Hamiltonian perturbation---causes the orbits to shrink in such a way that the orbital axis and eccentricity remain constant \cite{hamilton08}. Apparently, a~similar observation was made long ago in a somewhat different context by S.~Lie~\cite{ince,lie} for the Kepler problem.

\begin{figure}[t]
\centering
 \includegraphics[width=0.4\textwidth]{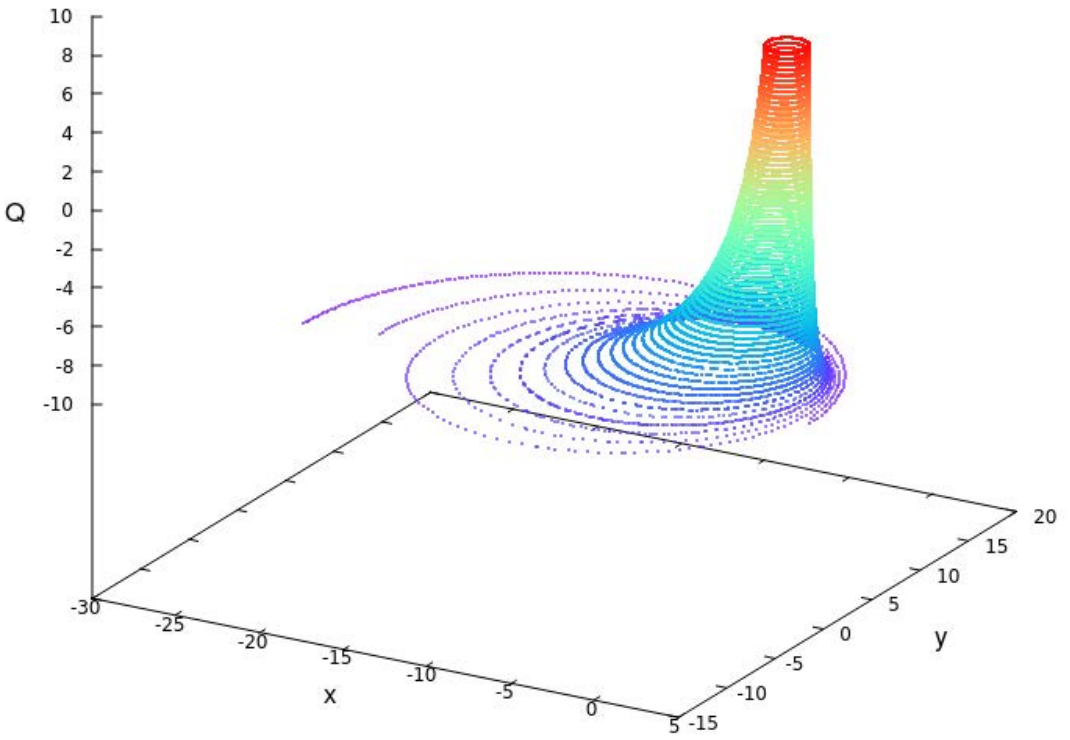}
 \qquad \includegraphics[width=0.4\textwidth]{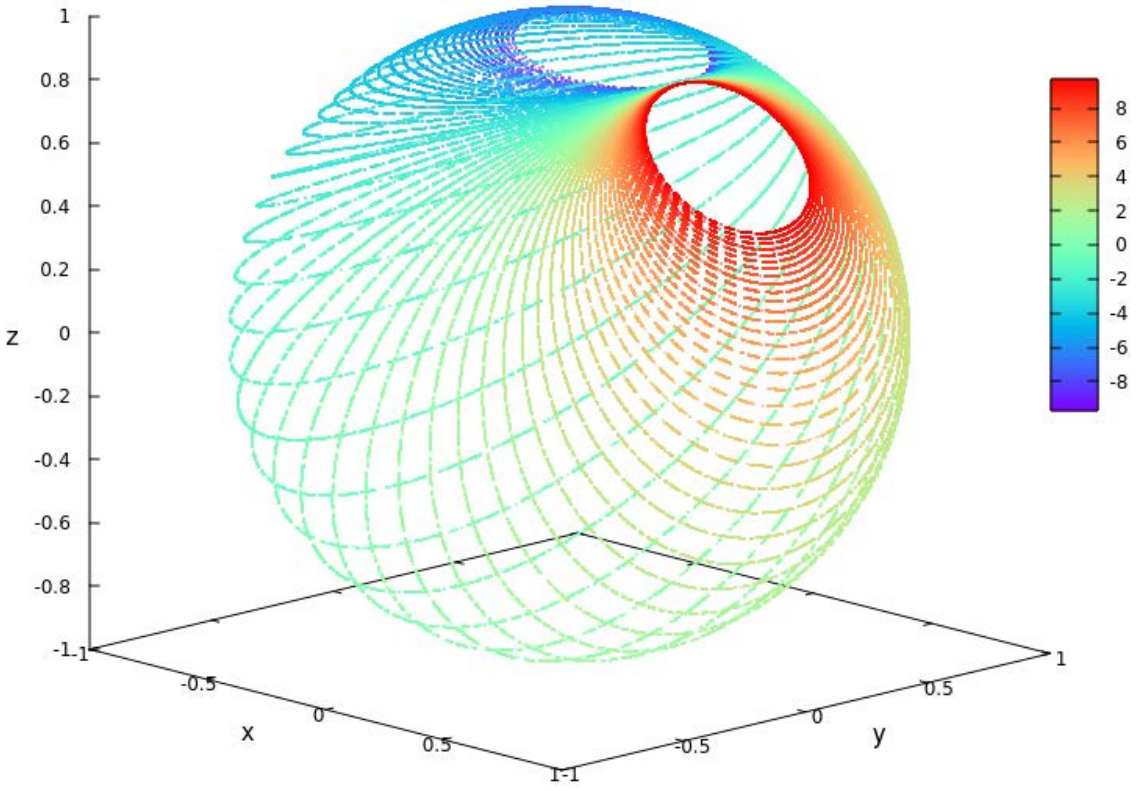}

\caption{Illustrating the change in the orbits on the (a) $xy$ plane (b) stereographic sphere caused by a~gradual change in the monopole charge $Q$ (indicated here by colors). In (a), every cross section parallel to the $xy$ plane (a specific value of $Q$) yields a circular orbit on the $xy$ plane. The orientation of the sphere in~(b) is the same as that of Figure~\ref{fig1}.}
\label{changingQ}
\end{figure}

In our problem, the perturbation (magnetic field) is Hamiltonian, but, as above, associated with forces linear in the velocity. As in the Kepler case, changing $Q$ will cause the constants associated with the motion to evolve; we now show that the rigidity of the Poisson algebra restricts the flow of the orbital parameters in a way strongly reminiscent of the ``Kepler with linear damping'' case; the orbits preserve their shape and orientation but the orbit's center moves. We also observe from direct integrations of the equations of motion that under a gradual change of $Q$, the centers of our circular orbits on the $xy$ plane move in a straight line, corresponding to a~shrinking of the length of the orbital axis while not disturbing the other constants of motion, just as in the Kepler case with linear damping (see Figure~\ref{changingQ}\,(a)).
Part of this phenomenology can be understood from the global structure of the space of orbits. Note that in the $Q=0$ case, since the negative of the potential is less than $1/r^2$ at large distances, the usual stability arguments indicate that there will be a range of angular momentum $L\in [0, L_{\max}]$ \big(with $L_{\max} = \sqrt{\frac{\alpha}{2{\cal R}^2}}$\big) that are associated with bound orbits. Note also that the effective potential in these cases has a positive value at its outermost local maximum, indicating the spectrum of finite orbits is an interval containing~0. Now including the effect of a magnetic field, the $Q$ shifts the $L_{\max}$ but the global picture remains the same. For one sign of $Q$ then an arbitrarily large value of $Q$ can be accommodated, whereas for the other sign of $Q$ there is a maximum, beyond which the orbital sense (i.e., sign of $L_z$) itself must flip to the other branch of the solutions (see Figure \ref{changingQ}\,(b)).

For \smash{$l^2 = \frac{\alpha}{2L_z^2} - {\cal R}^2 + \frac{Q}{2L_z}$}, where $l$ is the location of the centers of the orbits on the $xy$ plane, imposing the $l^2 > 0$ condition implies that $|Q/2L_z|<\alpha/2L_z^2 -{\cal R}^2$, i.e., $Q/L_z$ cannot be too negative for a fixed value of $L_z$. This allows us to plot Figures \ref{changingQ}\,(a) and \ref{changingQ}\,(b) with physical solutions of our system for both large and small values of $Q$ and $L_z$ which helps us further elucidate the global features of the space of orbits. For large $L_z$, the $Q\rightarrow\infty$ limit results in a~zero radius circle both on the $xy$ plane and on the stereographic sphere. Moreover, the gradual change in $Q$ allows us to see that the space of orbits forms something akin to the double covering of a 2-sphere because as we increase $Q$ in the direction where arbitrarily large values cannot be assigned, inverse projected trajectories on the sphere ``meet'' the oppositely oriented solution on the other side of the North pole on the sphere, i.e., the $L_z$ must change in sign (by having the orbits cross the North pole) such that more solutions on the sphere are obtainable, as described earlier.

Of historical note, the Newton off-center circular orbit potential with or without magnetic charge as fastened above admits a hodograph \cite{hamilton, maxwell} that is in a sense the inverse of that of the Kepler problem. The hodograph is a dynamically relevant version of the Gauss map of a~curve~\cite{gauss}. In the Gauss map, each point of a closed plane curve is mapped to the unit tangent vector there, the collection of which sweep out a complete circle some number of times. In the same sense, the hodograph of a closed orbit is the map from the points of the orbit to the velocity vector there; for the case of Kepler although the orbit is an ellipse, the velocity vectors sweep out a circle displaced from the origin (for a condensed yet easy-to-read rendition of the Kepler hodograph, see \cite{suzuki}). As expected, the `axis' of the off-center hodograph circle and the major axis of the elliptical orbit are normal.

In the case Newton off-center circular orbit potential, the $E=0$ orbits for any $Q$ are circles but their hodograph is an ellipse. One finds that the eccentricity of these `velocity vector ellipses' is $\epsilon=\frac{2Rl}{R^2+{\cal R}^2+ l^2}$ where $l$ is the distance of the center of the orbit from the origin and $R$ is the radius of the orbit in the plane, and that their axis is again normal to the `axis' of the orbit. Finally, note that the correspondence in both cases of conic sections only occur for the position and momenta; other higher order dynamical quantities of interest like acceleration, etc.\ do not form a conic section. This is presumably due to the fact that the algebra of the constants of motion we find in both Kepler and the off-center potential are linear in the momenta.

The existence of an everywhere attractive potential on the plane, leading to special classical zero energy orbits associated with a dynamical symmetry that survives the inclusion of magnetic flux of a magnetic monopole on the stereographic sphere is evocative of the Atiyah--Patodi--Singer index theorem in the quantum version of this problem \cite{bott, atiyah,singer}. In the interests of brevity of this note, our intention here is not to give a complete, detailed account of the quantum mechanical problem associated with the off-center circular orbit potential with a monopole field, but to just focus on a few phenomenological consequences of the dynamical symmetry.\looseness=1

Consider what happens to the spectrum of the model as we go from the classical to quantum version of this problem. In the $Q=0$ classical case, the spectrum consists of two overlapping continua, one that starts below zero and ends above zero (``bounded orbits'') and one that starts at zero and is unbounded positive (``unbound orbits''). Note that some of the bounded orbits are classically unstable. In the associated quantum problem, the spectrum always contains a~discrete spectrum of bound states for any $\alpha>0$, that terminates for generic values of $\alpha$ below~$E=0$, followed by a continuum of unbound states that starts at $E=0$ (Note this is easy to establish via variational considerations, and a physically and historically important example of the existence of bound states in two spatial dimensions for an everywhere attractive potential is \cite{cooper}). The continuity of the spectral flow in $\alpha$ indicates that at special values of $\alpha>0$ we should expect to get $E=0$ bound states.

For simplicity consider first the $Q=0$ solution of the associated quantum problem (here we take $m=1$ for notational convenience). Let the eigenvalue of the $L_z=I $. We have
\begin{equation}
\biggl[ -\frac{1}{2r}{\partial_r} r {\partial}_r + \frac{I^2}{2r^2} - \frac{\alpha}{\big(r^2+{\cal R}^2\big)^2} \biggr] \Psi = E\Psi,
\label{q1}
\end{equation}
which for $E=0$ and only when $\alpha = 2{\cal R}^2 I(I+1)$ admits the solution
\begin{equation}
\Psi_\pm(r) = A \frac {r^I{\rm e}^{\pm {\rm i}I\phi}}{\big(r^2+{\cal R}^2\big)^I}
\label{q2}
\end{equation}
from which we see that, on physical grounds, that $I$ must be integer (hermiticity of $L_z$) and~$I>0$ (normalizability). Note the familiar appearance of the analytic $z = r{\rm e}^{{\rm i}\phi}$ and the leading $z^I$ dependence of this solution, all strongly reminiscent of the analytic structure of the wave function in the integer quantum hall state \cite{haldane, laughlin} (for a recent treatment, see \cite{senthil}).

We anticipate the quantum version of the operators \eqref{J} will satisfy an operator algebra where the Poisson bracket in \eqref{Bnot0} is replaced by the commutator, becoming a state-classifying algebra in the $H=0$ subspace of which (for $Q=0$) \eqref{q2} is a member. Viewed from the stereographic sphere, there are no $E=0$ sections except constant functions, which lead to non-normalizable solutions on the plane (the conformal dimension of $\Psi$ is 0, and in two dimensions the kinetic energy, here the Laplace--Beltrami operator, is conformal dimension~$-2$). The~${\alpha = 2{\cal R}^2 I(I+1)}$ condition on the zero energy section indicates that the conformal transformation of the problem from the plane to the sphere is not $E^{S^2}=0$ but $E^{S^2} = {\rm const}$, as can also be seen from the quadratic Casimir element \eqref{Casimir}, which, as a consequence of requiring the commutator algebra to have a linear representation on a finite number of states, we have~${C_2 = I(I+1)}$, for some $I\in {\mathbb Z}$. Rewriting $\Psi_{\pm}$ in the $S^2$ coordinates reveals it is as the~$Y^{\pm I}_I (\theta, \phi)$ (in weight space notation, $|I,\pm I>$) state of spherical harmonics.

The operator algebra is universal to the sphere, and up to a shift in $L_z$ by a central element, is unchanged by the inclusion of monopole magnetic field. As noted long ago by Dirac \cite{dirac}, in the quantum problem requiring the wave functions to be single valued implies the quantization of the product, $M$, of the magnetic and electric charges; in our case this implies ($\hbar=1$) $M = \frac{Q}{2{\cal R}^2} \in {\mathbb Z}$. The quadratic Casimir element then indicates that there is a model (a particular value of $\alpha$) having $E=0$ states for each choice of $I$, $M$, though clearly not all choices of the later are possible. Let $m$ be the eigenvalue of the $L_z$ in the $Q\ne 0$ case; by hermiticity of $L_z$ we have~$m\in {\mathbb Z}$. Thus, the quantum version of \eqref{Casimir} indicates that $|m-M/2|$ is bounded by $I$.

It is now revealing to count the possible $E=0$ normalizable quantum solutions consistent with the operator algebra in this 2-d monopole background \cite{bardacki}. For example, at large $I$ the possible $|M|$ are limited to be less than $2I$ (since we require $\alpha >0$). We know that for $Q=0$ there are two zero modes on the plane if $\alpha$ is large enough; to simplify the discussion here we limit ourselves to that case in which in the absence of $Q$ there would be no zero modes, that is, we specialize to the largest possible value, $|M| = 2I$ (the smallest possible non-zero $\alpha$) and take~$M$ even and positive. By the forgoing constraint the possible $m$ values now range from 0 to $2I$. Since the $m=2I$ mode is not normalizable in the plane that leaves exactly $M$ zero modes in this case, as expected by the application of the index theorem. Technically, the index theorem counts the difference between zero modes of different chirality, but in two dimensions the chirality is manifest in the complex analytic factor in the wavefunction. In typical applications (as here) the zero modes are all of the same chirality; going to $M<0$ doesn't change the above counting argument but flips the chirality of the zero modes.

\section{Conclusion}
Newton's off-center circular orbit potential is ideal for a study at $E=0$ that includes dynamical symmetry, their consequence for perturbation linear in the velocity, hodography and the index theorem of elliptic operators. In this brief note we've sought to display some aspects of these connections with the dynamical symmetry and kinematic consequence of the associated situation in the Kepler problem. The $E=0$ problem has a conformal image as motion on a two sphere that containing a magnetic monopole at its center. The confluence of the rigidity of the dynamical algebra and the topology of the monopole field leads to a simple counting of possible quantum solutions with $E=0$, the `zero modes', that can be understood as a straightforward consequence of the simplest version of the index theorem.

\subsection*{Acknowledgements}

The authors acknowledge partial support via NSF grant DMR-2226956. We thank Don Priour for conversations and assistance with the figures. We are thankful to a referee for bringing reference \cite{cisneros} to our attention.

\pdfbookmark[1]{References}{ref}
\LastPageEnding

\end{document}